\documentclass[10pt,conference,letterpaper]{IEEEtran}

\usepackage{amsmath}
\usepackage{amsfonts}
\usepackage{amssymb,bm}
\usepackage{amsthm}
\usepackage{algorithm}
\usepackage{algpseudocode}
\usepackage{cite}
\usepackage{graphicx}
\usepackage{epstopdf}
\usepackage[caption=false,font=footnotesize]{subfig}
\usepackage{fixltx2e}
\usepackage{balance}
\usepackage{cases}

\IEEEoverridecommandlockouts
\begin{document}
%
\title{A Rate-Splitting Strategy for Max-Min Fair Multigroup Multicasting}
\author{
\IEEEauthorblockN{Hamdi Joudeh\IEEEauthorrefmark{1} and Bruno Clerckx\IEEEauthorrefmark{1}\IEEEauthorrefmark{2}}
\fontsize{9}{9}\upshape
\IEEEauthorrefmark{1} Department of Electrical and Electronic Engineering, Imperial College London, United Kingdom \\
\IEEEauthorrefmark{2} School of Electrical Engineering, Korea University, Seoul, Korea \\
\fontsize{9}{9}\selectfont\ttfamily\upshape
Email: \{hamdi.joudeh10, b.clerckx\}@imperial.ac.uk
\thanks{This work has been partially supported by the UK EPSRC under grant number EP/N015312/1.}
\thanks{978-1-5090-1749-2/16/\$31.00 \copyright2016 IEEE}
}
\maketitle

\begin{abstract}
%
We consider the problem of transmit beamforming to multiple cochannel multicast groups.
The conventional approach is to beamform a designated data stream to each group, while treating potential inter-group interference as noise at the receivers.
In overloaded systems where the number of transmit antennas is insufficient to perform interference nulling,
we show that inter-group interference dominates at high SNRs, leading to a saturating max-min fair performance.
We propose a rather unconventional approach to cope with this issue based on the concept of Rate-Splitting (RS). In particular, part of the interference is broadcasted to all groups such that it is decoded and canceled before the designated beams are decoded.
We show that the RS strategy achieves significant performance gains over the conventional multigroup multicast beamforming strategy.
\end{abstract}


\IEEEpeerreviewmaketitle
\section{Introduction}
\label{Section_Introduction}
\newcounter{Proposition_Counter} 
\newcounter{Theorem_Counter} 
\newcounter{Lemma_Counter} 
\newcounter{Remark_Counter} 
\newcounter{Assumption_Counter}
\newcounter{Definition_Counter}
Since the work of \emph{Sidiropoulos et al.} \cite{Sidiropoulos2006}, beamforming for physical-layer multicasting has received
considerable research attention.
In the most basic setup, the Base Station (BS) transmits a common data stream to all receivers.
This was later generalized to multiple cochannel multicast groups, also known as multigroup multicasting \cite{Karipidis2008}.
The main problems considered in the multicasting literature are those of classical multiuser beamforming, namely the Quality of Service (QoS) constrained power minimization problem and the power constrained Max-Min Fair (MMF) problem.
Such problems were shown to be NP-hard, and the solutions advised in \cite{Sidiropoulos2006,Karipidis2008} are based on Semidefinite Relaxation (SDR) and Gaussian randomization techniques.
Alternative solutions based on convex approximation methods were later proposed, exhibiting marginally improved performances under certain setups, and more importantly, lower complexities \cite{Bornhorst2011,Schad2012}.
The multigroup multicasting problem was also extended to incorporate per-antenna power constraints \cite{Christopoulos2014} and large-scale arrays \cite{Christopoulos2015}.
In addition to the QoS and MMF problems, the sum-rate maximization problem was considered in \cite{Christopoulos2014a}.
The common transmission strategy adopted in multigroup multicasting is based on extending the multiuser beamforming paradigm, i.e. each message is first encoded into an independent data stream then transmitted through linear precoding (or beamforming).
However, the multicast nature of each stream results in different, and more difficult, design problems compared to their multiuser counterparts.
In the beamforming strategy, each receiver decodes its desired stream while treating all interfering streams as noise.
Hence, inter-group interference is inevitable under an insufficient number of BS antennas.
Although rarely highlighted or treated in the multigroup multicasting literature, such interference can be highly detrimental.

We propose a beamforming strategy based on the concept of Rate-Splitting (RS), where the message intended to each group is split into a common part and a designated part.
All common parts are packed into one super common message, broadcasted to all users in the system.
Designated parts on the other hand are transmitted in the conventional beamforming manner.
While the concept of RS is not particularly new (it appears in the interference channel literature),
it has only been applied recently to multiuser beamforming, where it was shown to enhance the performance under residual interference arising from imperfect Channel State Information (CSI) at the BS \cite{Joudeh2016a,Joudeh2016b}.
We show that RS brings significant performance gains to multigroup multicasting, particularly in inter-group interference limited scenarios.
While the focus is on the MMF problem in this paper, RS can be extended to the QoS problem.

The rest of the paper is organized as follows. Section \ref{Section_System_Model} presents the system model.
The limitations of the conventional transmission strategy are analysed in Section \ref{Section_NoRS_Problem}.
In Section \ref{Section_RS}, the RS strategy is introduced and the performance gains over the conventional strategy are derived.
The RS precoders are optimized using a Weighted Minimum Mean Square Error (WMMSE) algorithm in Section \ref{Section_Precoder_Opt}.
Simulation results are presented in Section \ref{Section_Simulations}, and Section \ref{Section_Conclusion} concludes the paper.
\section{System Model}
\label{Section_System_Model}
Consider a transmitter equipped with $N$ antennas communicating with $K$ single-antenna receivers
grouped into the $M$ multicast groups $\{\mathcal{G}_{1},\ldots,\mathcal{G}_{M}\}$, where $\mathcal{K} \triangleq \{ 1, \ldots, K \}$
and
$\mathcal{M} \triangleq \{ 1, \ldots, M\}$.
We assume that $\bigcup_{m\in\mathcal{M}} \mathcal{G}_{m} = \mathcal{K}$,
and $\mathcal{G}_{m} \cap \mathcal{G}_{j} =\emptyset$, for all $m,j \in \mathcal{M}$ and
$m\neq j$.
Let $\mathbf{x} \in \mathbb{C}^{N}$ denote the signal vector transmitted by the BS in a given channel use, which  is subject to an average power constraint $\mathrm{E}\left\{ \mathbf{x}^{H}\mathbf{x} \right\} \leq P$.
Denoting the corresponding signal received by the $k$th user as $y_{k}$, the input-output relationship writes as
$y_{k} = \mathbf{h}_{k}^{H} \mathbf{x} + n_{k}$,
where $\mathbf{h}_{k}\in \mathbb{C}^{N}$ is the narrow-band channel vector from the BS to the $k$th user,
and $n_{k} \in \mathcal{CN}(0,\sigma_{\mathrm{n},k}^{2})$ is the Additive White Gaussian Noise (AWGN) at the receiver.
We assume, without loss of generality, that $\sigma_{\mathrm{n},1}^{2},\ldots,\sigma_{\mathrm{n},K}^{2} =
\sigma_{\mathrm{n}}^{2}$, from which the transmit SNR is given by $P/\sigma_{\mathrm{n}}^{2}$.
Moreover, the transmitter perfectly knows all $K$ channel vectors, and each receiver knows its own channel vector.
In multigroup multicast transmission, the BS communicates the messages
$W_{1},\ldots,W_{M}$ to $\mathcal{G}_{1},\ldots,\mathcal{G}_{M}$ respectively.
Consider a conventional linear precoding (beamforming) transmission model.
Messages are first encoded into independent data streams, where the vector of coded data symbols in a given channel use
writes as $\mathbf{s}_{\mathrm{p}} \triangleq [s_{1},\ldots,s_{M}]^{T} \in \mathbb{C}^{M}$.
We assume that $\mathrm{E}\left\{\mathbf{s}_{\mathrm{p}}\mathbf{s}_{\mathrm{p}}^{H} \right\} = \mathbf{I}$, where power allocation is considered  part of the beamforming.
Data streams are then mapped to the transmit antennas through a linear precoding matrix $\mathbf{P}_{\mathrm{p}} \triangleq [\mathbf{p}_{1},\ldots,\mathbf{p}_{M}]$, where $\mathbf{p}_{m} \in \mathbb{C}^{N}$ is the $m$th group's precoding  vector.
The resulting transmit signal is
\begin{equation}
\label{Eq_x_NoRS}
\mathbf{x} =  \sum_{m=1}^{M} \mathbf{p}_{m}s_{m}
\end{equation}
where the power constraint reduces to $\sum_{m=1}^{M} \|\mathbf{p}_{m}\|^{2} \leq P$.
The $k$th user's average receive power (over multiple channel uses in which the channel is fixed)  writes as
\begin{equation}
\label{Eq_T_k}
T_{k} = \overbrace{|\mathbf{h}_{k}^{H}\mathbf{p}_{\mu(k)}|^{2}}^{S_{k}} + \overbrace{\sum_{m\neq \mu(k)} |\mathbf{h}_{k}^{H}\mathbf{p}_{m}|^{2} + \sigma_{\mathrm{n}}^{2}}^{I_{k}}.
\end{equation}
where $\mu: \mathcal{K} \mapsto \mathcal{M}$  maps a user-index to the corresponding group-index, i.e. $\mu(k) = m$ such that $k \in \mathcal{G}_{m}$. In the following, $\mu(k)$ is referred to as $\mu$ for brevity where the argument of the function is clear from the context.
In \eqref{Eq_T_k}, $S_{k}$ and $I_{k}$ denote the desired receive power and the interference plus noise power, respectively.
Hence, the Signal to Interference plus Noise Ratio (SINR) experienced by the $k$th user is defined as $\gamma_{k} \triangleq S_{k}I_{k}^{-1}$.
Under Gaussian signalling, the $k$th achievable user-rate is given by $R_{k} = \log_{2}(1 + \gamma_{k} )$.
In multigroup multicasting, users belonging to the same group decode the same data stream.
Therefore, to guarantee that all users in the $m$th group are able to recover $W_{m}$ successfully, the corresponding code-rate should not exceed the group-rate defined as $r_{m} \triangleq \min_{i \in \mathcal{G}_{m}} R_{i}$.
\section{Max-Min Fairness and Inter-Group Interference}
\label{Section_NoRS_Problem}
In the light of the conventional multi-stream beamforming model, the MMF problem is formulated as
\begin{equation}
\label{Eq_Problem_R_NoRS}
\mathcal{R}(P):
\begin{cases}
       \underset{\mathbf{P}_{\mathrm{p}}}{\max} & \underset{m \in \mathcal{M}}{\min} \
        \underset{i \in \mathcal{G}_{m}}{\min} \ R_{i}  \\
       \text{s.t.}  & \displaystyle{\sum_{m=1}^{M} \|\mathbf{p}_{m}\|^{2} \leq P}
\end{cases}
\end{equation}
where the inner minimization in \eqref{Eq_Problem_R_NoRS} accounts for the multicast nature within each group, while  the outer minimization accounts for the fairness across groups.
It is common practice to formulate the above problem in terms of the SINRs \cite{Karipidis2008,Schad2012,Christopoulos2014,Christopoulos2015}.
Since each group receives a single stream, and due to the Rate-SINR monotonic relationship, the two formulations are equivalent.
The rate formulation is preferred in this work in order to compare the performance to the RS scheme.
\subsection{Inter-Group Interference and Degrees of Freedom}
\label{Subsection_DoF_definition}
An optimum MMF design achieves balanced group rates, requiring a simultaneous increase in powers allocated to all streams as $P$ increases.
In scenarios where the number of transmit antennas in insufficient to place each beam in the null space of all its unintended groups, inter-group interference is expected to limit the MMF performance.
To characterize this, we resort to high SNR analysis through the Degrees of Freedom (DoF).
This regime is of particular interest as the effect of noise can be neglected, and inter-group interference is the main limiting factor.
The DoF can be roughly interpreted as the number of interference-free streams that can be simultaneously communicated in a single channel use.
To facilitate the definition of the DoF, we first define a precoding scheme $\left\{ \mathbf{P}_{\mathrm{p}} (P) \right\}_{P}$ as a family of feasible precoders with one precoding matrix for each power level.
The corresponding achievable user-rates write as
$\left\{R_{1}(P),\ldots,R_{K}(P)\right\}_{P}$, and the $k$th user-DoF is defined as $D_{k} \triangleq \lim_{P \rightarrow \infty} \frac{R_{k}(P) }{\log_{2}(P)}$.
It follows that the $m$th group-DoF, denoted by $d_{m}$, satisfies
$ 0 \leq d_{m}  \leq D_{i}$ for all $i \in \mathcal{G}_{m}$.
The corresponding symmetric-DoF is given by  $d = \min_{m\in \mathcal{M}}d_{m}$.

For a given setup, the optimum MMF precoding scheme is denoted by $\left\{ \mathbf{P}_{\mathrm{p}}^{\ast} (P) \right\}_{P}$.
The corresponding MMF-DoF is given by $d^{\ast} = \lim_{P \rightarrow \infty}  \frac{\mathcal{R}(P)}{\log_{2}(P)}$, which is the maximum symmetric-DoF.
Since each user is equipped with a single antenna, then $D_{1},\ldots,D_{K} \leq 1$, and $d \leq 1$ for any precoding scheme.
Hence, when $d=1$ is achievable, it is also optimum.
It should be noted that although a rate-optimal precoder is also optimum in a DoF sense, the converse is usually untrue.
%
\subsection{MMF-DoF of Multi-Stream Beamforming}
In the DoF analysis, we make the following assumptions.
\newtheorem{Assumption_Generic_Channel}[Assumption_Counter]{Assumption}
\begin{Assumption_Generic_Channel}
\label{Assumption_Generic_Channel}
\textnormal{
The channel vectors $\mathbf{h}_{1},\ldots,\mathbf{h}_{K}$ are independently drawn from a set of continuous distribution functions.
Hence, for any $N\times K_{\mathrm{sub}}$ matrix in which the $K_{\mathrm{sub}}$ column vectors constitute any subset of the $K$ channel vectors, it holds with probability one that the rank is $\min\{ N,K_{\mathrm{sub}} \}$.
}
\end{Assumption_Generic_Channel}
\newtheorem{Assumption_Equal_Groups}[Assumption_Counter]{Assumption}
\begin{Assumption_Equal_Groups}
\label{Assumption_Equal_Groups}
\textnormal{
We assume equal size groups. i.e. $|\mathcal{G}_{1}|,\ldots,|\mathcal{G}_{M}| = G$, where $G = K/M$ is a positive integer.
}
\end{Assumption_Equal_Groups}
Next, the MMF-DoF of the conventional multi-stream transmission scheme is characterized.
\newtheorem{Proposition_DoF_NoRS}[Proposition_Counter]{Proposition}
\begin{Proposition_DoF_NoRS}\label{Proposition_DoF_NoRS}
\textnormal{
Under Assumptions \ref{Assumption_Generic_Channel} and \ref{Assumption_Equal_Groups},
the optimum MMF-DoF achieved by solving \eqref{Eq_Problem_R_NoRS} is given by
\begin{align}
\label{Eq_max_min_DoF_NoRS}
\lim_{P \rightarrow \infty}  \frac{\mathcal{R}(P)}{\log_{2}(P)}=
\begin{cases}
       1, &  N \geq  N_{\min} \\
       0, &   N  < N_{\min}
\end{cases}
\end{align}
where $N_{\min} = 1 + K - G$.
}
\end{Proposition_DoF_NoRS}
To show this, let us define $\mathbf{H}_{m}$ as the matrix with columns constituting channel vectors of all users in $\mathcal{G}_{m}$, and
$\bar{\mathbf{H}}_{m} = [\mathbf{H}_{1},\ldots,\mathbf{H}_{m-1},\mathbf{H}_{m+1},\ldots,\mathbf{H}_{M}]$ as the complementary set of channel vectors.
By Assumptions \ref{Assumption_Generic_Channel} and \ref{Assumption_Equal_Groups}, $\mathrm{null}\big( \bar{\mathbf{H}}_{m}  \big)$ has a  dimension of $\max \{N+G-K,0\}$ for all $m \in \mathcal{M}$.
Hence, $ N \geq  N_{\min}$ is sufficient to place each beamforming vector in the null space of all groups it is not intended to, i.e. $\mathbf{p}_{m} \in \mathrm{null}\big( \bar{\mathbf{H}}_{m}  \big)$ for all $m \in \mathcal{M}$.
Each group sees no inter-group interference, and a DoF of $1$ per group is achievable.
Such DoF is optimum as it cannot be surpassed.
On the contrary, when $N  < N_{\min}$, this is not possible, and inter-group interference limits the MMF-DoF to $0$ as shown in the Appendix.
We refer to this case as an overloaded system.

Finally, we conclude this section by highlighting the impact of a collapsing DoF on the rate performance.
When $d = 0$, the MMF rate stops growing as SNR grows large, reaching a saturated performance\footnote{To be more precise, this corresponds to a rate scaling as $o\big(\log_{2}(P)\big)$, which either stops growing or grows extremely slow with $P$ compared to the interference free scenario, reaching a flat or almost-flat performance.}.
Although the DoF analysis is carried out as SNR goes to infinity, its results are highly visible in finite SNR regimes as we see in the simulation results.
\section{Rate-Splitting For Multigroup Multicasting}
\label{Section_RS}
The saturating performance can be avoided by single-stream multigroup transmission.
In particular, the $M$ messages are packed into one super message, encoded into a single data stream.
This is broadcasted such that it is decoded by all groups, hence retrieving their corresponding messages.
Since this interference-free transmission achieves a total DoF of $1$, each group is guaranteed a non-saturating performance with a DoF of $1/M$.
However, relying solely on this strategy jeopardizes partial gains potentially achieved by multi-stream beamforming.
A simple example is the low-SNR regime, where interference is overwhelmed by noise, and beamforming each message to its corresponding group is a preferred strategy.
Hence, we introduce the following unifying strategy.
\subsection{The Rate-Splitting Strategy}
Each group-message is split into a common part and a group-designated part, e.g. $W_{m} = \{W_{m0},W_{m1}\}$, with $W_{m0}$ and $W_{m1}$ being the common and designated parts respectively.
All common parts are packed into one super common message $W_{\mathrm{c}} \triangleq \{W_{10},\ldots,W_{M0}\}$,
encoded into the stream $s_{\mathrm{c}}$, and then precoded using $\mathbf{p}_{\mathrm{c}} \in \mathbb{C}^{N}$.
On the other hand, the designated messages are encoded into $s_{1},\ldots,s_{M}$ and precoded in the conventional multi-stream manner described in Section \ref{Section_System_Model}.
The transmit signal writes as a superposition of the common stream and the designated streams such that
\begin{equation}
\label{Eq_x_RS}
\mathbf{x} =  \mathbf{p}_{\mathrm{c}} s_{\mathrm{c}} + \sum_{m=1}^{M} \mathbf{p}_{m} s_{m}.
\end{equation}
The power constraint writes as $\|\mathbf{p}_{\mathrm{c}}\|^{2} + \sum_{m=1}^{M} \|\mathbf{p}_{m}\|^{2} \leq P$.
The common stream can be interpreted as the part of the interference that is decoded (hence eliminated) by all groups,
while interference from designated streams is treated as noise.
The $k$th user's average received power now writes as
\begin{equation}
\label{Eq_T_c_k}
T_{\mathrm{c},k} = |\mathbf{h}_{k}^{H}\mathbf{p}_{\mathrm{c}}|^{2} + T_{k}
\end{equation}
where $S_{\mathrm{c},k} = |\mathbf{h}_{k}^{H}\mathbf{p}_{\mathrm{c}}|^{2}$ denotes the common stream's receive power
and $I_{\mathrm{c},k} = T_{k}$ is the interference plus noise power experienced by the common stream.
By treating all designated streams as noise, the SINR of the common stream at the $k$th receiver is given by
$\gamma_{\mathrm{c},k} \triangleq S_{\mathrm{c},k}I_{\mathrm{c},k}^{-1}$.
Hence, transmitting $W_{\mathrm{c}}$ at a rate of $R_{\mathrm{c},k} = \log_{2}(1 + \gamma_{\mathrm{c},k} )$
guarantees successful decoding by the $k$th receiver.
To guarantee that $W_{\mathrm{c}}$ is successfully recovered by all receivers, the rate of the common data stream should not exceed the common-rate defined as
$R_{\mathrm{c}} = \min_{k\in \mathcal{K}}  R_{\mathrm{c},k}$.
After decoding the common stream, the receiver removes it from $y_{k}$ using Successive Interference Cancellation (SIC).
This is followed by decoding the designated stream in the presence of the remaining interference and noise,
achieving the rate $R_{k}$ defined in Section \ref{Section_System_Model}.
The common rate writes as a sum of $M$ portions: $R_{\mathrm{c}} = \sum_{m=1}^{M} C_{m}$,
where $C_{m}$ is associated with $W_{m0}$.
It follows that the $m$th group-rate is defined as $R_{\mathrm{g},m} \triangleq C_{m} + \min_{i \in \mathcal{G}_{m}}R_{i}$,
consisting of a common-rate portion plus a designated-rate.
It is easy to see that the RS group-rates reduce to the conventional group-rates defined in Section \ref{Section_NoRS_Problem} when $|W_{\mathrm{c}}| = 0$.

\subsection{MMF With Rate-Splitting}
The MMF problem is formulated in terms of RS as follows
\begin{equation}
\label{Eq_Problem_R_RS}
\mathcal{R}_{\mathrm{RS}}(P):
\begin{cases}
       \underset{\mathbf{c},\mathbf{P}}{\max} & \underset{m \in \mathcal{M}}{\min}
       \Big( C_{m} + \underset{i \in \mathcal{G}_{m}}{\min} \ R_{i} \Big) \\
       \text{s.t.}
       &  R_{\mathrm{c},k} \geq \sum_{m=1}^{M} C_{m} , \forall k \in \mathcal{K}   \\
       &   C_{m} \geq 0, \forall m \in \mathcal{M} \\
       & \displaystyle{ \|\mathbf{p}_{\mathrm{c}}\|^{2} +  \sum_{m=1}^{M} \|\mathbf{p}_{m}\|^{2} \leq P}
\end{cases}
\end{equation}
where $\mathbf{c} \triangleq [C_{1},\ldots,C_{M}]^{T}$ is the vector of common-rate portions.
The first set of constraints in \eqref{Eq_Problem_R_RS} accounts for the global multicast nature of the common stream and guarantees that it can be decoded by all users.
The second set of constraints guarantees that no user is allocated a negative common-rate portion.
Solving \eqref{Eq_Problem_R_RS} yields the optimum precoding matrix, in addition to the splitting ratio for each group-message.
Next, the DoF performance of the RS scheme is characterized.
\newtheorem{Proposition_DoF_RS}[Proposition_Counter]{Proposition}
\begin{Proposition_DoF_RS}\label{Proposition_DoF_RS}
\textnormal{
The MMF-DoF achieved by solving the RS problem in \eqref{Eq_Problem_R_RS} is lower-bounded by
\begin{align}
\label{Eq_max_min_DoF_RS}
\lim_{P \rightarrow \infty}  \frac{\mathcal{R}_{\mathrm{RS}}(P)}{\log_{2}(P)} \geq
\begin{cases}
       1, &  N \geq  N_{\min} \\
       \frac{1}{M}, &   N  < N_{\min}
\end{cases}
\end{align}
where the lower-bound is tight for $N \geq  N_{\min}$.}
\end{Proposition_DoF_RS}
This follows directly from Proposition \ref{Proposition_DoF_NoRS}, and the fact that the single-stream multigroup solution described at the beginning of this section is feasible for problem \eqref{Eq_Problem_R_RS}.
\section{Precoder Optimization}
\label{Section_Precoder_Opt}
In the RS scheme, each group-rate writes as a sum of two rate components.
Hence, the MMF solutions in \cite{Karipidis2008,Schad2012} do not apply here, as the performance metric of each user cannot be expressed as a single SINR.
Alternatively, we resort to the WMMSE approach  \cite{Christensen2008,Shi2011}, which is particularly effective in dealing with problems incorporating non-convex coupled sum-rate expressions, including RS problems \cite{Joudeh2016a,Joudeh2016b}.
\subsection{Rate-WMMSE Relationship}
We start by defining the MSEs.
The $k$th user's estimate of $s_{\mathrm{c}}$, denoted by $\widehat{s}_{\mathrm{c},k}$,  is obtained by applying the  equalizer $g_{\mathrm{c},k}$ to the receive signal such that $\widehat{s}_{\mathrm{c},k} = g_{\mathrm{c},k}y_{k}$.
After removing the common stream using SIC, the equalizer $g_{k}$ is applied to the remaining signal to obtain an estimate of $\widehat{s}_{k}$ given by  $\widehat{s}_{k}=g_{k}(y_{k}-\mathbf{h}_{k}^{H}\mathbf{p}_{\mathrm{c}}s_{\mathrm{c}})$.
The common and private MSEs at the output of the $k$th receiver, defined as $\varepsilon_{\mathrm{c},k} \triangleq \mathrm{E}\{|\widehat{s}_{\mathrm{c},k} - s_{\mathrm{c}}|^{2}\}$ and $\varepsilon_{k} \triangleq \mathrm{E}\{|\widehat{s}_{k} - s_{k}|^{2}\}$ respectively, write as:
\begin{subequations}
\label{Eq_MSE}
\begin{align}
  \label{Eq_MSE_c_k}
  \varepsilon_{\mathrm{c},k} & = |g_{\mathrm{c},k}|^{2} T_{\mathrm{c},k} -2\Re \big\{g_{\mathrm{c},k}\mathbf{h}_{k}^{H}\mathbf{p}_{\mathrm{c}}\big\}+1 \\
  \label{Eq_MSE_k}
  \varepsilon_{k} & =  |g_{k}|^{2} T_{k}-2\Re \big\{g_{k}\mathbf{h}_{k}^{H}\mathbf{p}_{\mu}\big\}+1.
\end{align}
\end{subequations}
The MMSEs are defined as
$\varepsilon_{\mathrm{c},k}^{\mathrm{MMSE}}  \triangleq \min_{g_{\mathrm{c},k}} \varepsilon_{\mathrm{c},k} =  T_{\mathrm{c},k}^{-1} I_{\mathrm{c},k} $
and
$\varepsilon_{k}^{\mathrm{MMSE}}  \triangleq \min_{g_{k}} \varepsilon_{k} =
T_{k}^{-1}I_{k}$,
where the corresponding optimum equalizers are the well-known MMSE weights written as
$g_{\mathrm{c},k}^{\mathrm{MMSE}} = \mathbf{p}_{\mathrm{c}}^{H}\mathbf{h}_{k} T_{\mathrm{c},k}^{-1}$
and
 $g_{k}^{\mathrm{MMSE}} = \mathbf{p}_{k}^{H}\mathbf{h}_{k}T_{k}^{-1}$.
The MMSEs are related to the SINRs such that
$\gamma_{\mathrm{c},k} = \big( 1/\varepsilon_{\mathrm{c},k}^{\mathrm{MMSE}} \big) - 1 $
and
$\gamma_{k} = \big( 1/\varepsilon_{k}^{\mathrm{MMSE}} \big) - 1 $,
from which the achievable rates write as
$R_{\mathrm{c},k} = -\log_{2}(\varepsilon_{\mathrm{c},k}^{\mathrm{MMSE}})$ and
$R_{k} = -\log_{2}(\varepsilon_{k}^{\mathrm{MMSE}})$.

Next we introduce the main building blocks of the solution, the augmented WMSEs defined for the $k$th user as:
\begin{equation}
\label{Eq_A_WMSEs}
\xi_{\mathrm{c},k}
 \triangleq
u_{\mathrm{c},k} \varepsilon_{\mathrm{c},k}  -  \log_{2} (  u_{\mathrm{c},k} )
\ \ \ \text{and} \ \ \
\xi_{k}
 \triangleq
u_{k} \varepsilon_{k}  -  \log_{2}  (  u_{k} )
\end{equation}
where $u_{\mathrm{c},k}, u_{k} > 0$ are the corresponding weights.
In the following, $\xi_{\mathrm{c},k}$ and $\xi_{k}$ are referred to as the WMSEs for brevity.
The Rate-WMMSE relationship is established by optimizing \eqref{Eq_A_WMSEs} over the equalizers and weights such that:
\begin{subequations}
\label{Eq_min_WMSE}
\begin{align}
\label{Eq_min_WMSE_c}
 \xi_{\mathrm{c},k}^{\mathrm{MMSE}} & \triangleq  \underset{u_{\mathrm{c},k}, g_{\mathrm{c},k}}{\min} \xi_{\mathrm{c},k} = 1-R_{\mathrm{c},k}
\\
\label{Eq_min_WMSE_k}
 \xi_{k}^{\mathrm{MMSE}} & \triangleq \underset{u_{k}, g_{k}}{\min} \ \xi_{k}= 1-R_{k}
\end{align}
\end{subequations}
where the optimum equalizers are given by: $g_{\mathrm{c},k}^{\ast}  = g_{\mathrm{c},k}^{\mathrm{MMSE}} $ and $g_{k}^{\ast} = g_{k}^{\mathrm{MMSE}}$, and the optimum weights are given by:
$u_{\mathrm{c},k}^{\ast} = u_{\mathrm{c},k}^{\mathrm{MMSE}} \triangleq \big( \varepsilon_{\mathrm{c},k}^{\mathrm{MMSE}} \big)^{-1}$
and
$u_{k}^{\ast} = u_{k}^{\mathrm{MMSE}} \triangleq \big( \varepsilon_{k}^{\mathrm{MMSE}} \big)^{-1}$,
obtained by checking the first order optimality conditions.
By closely examining each WMSE, it can be seen that it is convex in each variable while fixing the other two.
\subsection{WMSE Reformulation and Algorithm}
Motivated by the relationship in \eqref{Eq_min_WMSE}, an equivalent WMSE reformulation of problem \eqref{Eq_Problem_R_RS} writes as
\begin{equation}
\label{Eq_Problem_WMSE_RS}
\widehat{\mathcal{R}}_{\mathrm{RS}}(P):
\begin{cases}
       \underset{r_{\mathrm{g}},\mathbf{r}, \mathbf{c},\mathbf{P},\mathbf{g},\mathbf{u}}{\max} \ \  r_{\mathrm{g}} \\
        \text{s.t.} \quad
         C_{m} + r_{m} \geq  r_{\mathrm{g}}, \forall m \in \mathcal{M}  \\
        \quad \quad 1-\xi_{i} \geq r_{m}, \forall i \in \mathcal{G}_{m}, \forall m \in \mathcal{M} \\
        \quad \quad 1-\xi_{\mathrm{c},k} \geq \sum_{m=1}^{M} C_{m} , \forall k \in \mathcal{K}   \\
        \quad \quad  C_{m} \geq 0, \forall m \in \mathcal{M} \\
        \quad \quad \displaystyle{ \|\mathbf{p}_{\mathrm{c}}\|^{2} +  \sum_{m=1}^{M} \|\mathbf{p}_{m}\|^{2} \leq P}
\end{cases}
\end{equation}
where $r_{\mathrm{g}}$ and $\mathbf{r} \triangleq \{r_{1},\ldots,r_{M}\}$  are auxiliary variables,
$\mathbf{u} \triangleq \{ u_{\mathrm{c},k},u_{k} \mid k \in \mathcal{K} \}$ is is the set of weights, and $\mathbf{g} \triangleq \{ g_{\mathrm{c},k},g_{k}\mid k \in \mathcal{K} \}$ is the set of equalizers.
The equivalence between \eqref{Eq_Problem_WMSE_RS} and \eqref{Eq_Problem_R_RS} is established by observing that the WMSEs are decoupled in their equalizers and weights. Hence, optimum $\mathbf{g}$ and $\mathbf{u}$ are obtained by minimizing each WMSE separately as shown in \eqref{Eq_min_WMSE}, yielding the MMSE solution.
The equivalence follows by substituting \eqref{Eq_min_WMSE} into \eqref{Eq_Problem_WMSE_RS}.
The WMSE problem in \eqref{Eq_Problem_WMSE_RS} is solved using an Alternating Optimization (AO) algorithm, which exploits its block-wise convexity.
In a given iteration of the algorithm, $\mathbf{g}$ and $\mathbf{u}$ are firstly updated using the optimum MMSE solution of \eqref{Eq_min_WMSE}.
Next, the set of precoders $\mathbf{P}$ alongside all auxiliary variables in \eqref{Eq_Problem_WMSE_RS}
are updated by solving $\widehat{\mathcal{R}}_{\mathrm{RS}}^{\mathrm{MMSE}}(P)$,
formulated by fixing $\mathbf{g}$ and $\mathbf{u}$ in \eqref{Eq_Problem_WMSE_RS}.
This is a convex problem which can be efficiently solved using interior-point methods \cite{Boyd2004}. The steps of the AO procedure are summarized in Algorithm \ref{Algthm_AO}.
\vspace{-1.5mm}
\begin{algorithm}
\caption{Alternating Optimization}
\label{Algthm_AO}
\begin{algorithmic}[1]
\State Initialize: $n\gets 0$, $\mathbf{P}$, $r_{\mathrm{g}}^{(n)} \gets 0$
\Repeat
    \State $n\gets n+1$
    \State $\big(g_{\mathrm{c},k},g_{k}\big)
    \gets \big(g_{\mathrm{c},k}^{\mathrm{MMSE}},g_{k}^{\mathrm{MMSE}}\big)$,
    $\forall k \in \mathcal{K}$
    \State $\big(u_{\mathrm{c},k},u_{k}\big)
    \gets \big(u_{\mathrm{c},k}^{\mathrm{MMSE}},u_{k}^{\mathrm{MMSE}}\big)$,
    $\forall k \in \mathcal{K}$
    \State $(r_{\mathrm{g}}^{(n)},\mathbf{r},\mathbf{c},\mathbf{P}) \gets  \arg{\widehat{\mathcal{R}}_{\mathrm{RS}}^{\mathrm{MMSE}}(P)}$
\Until$|r_{\mathrm{g}}^{(n)} - r_{\mathrm{g}}^{(n-1)} \big| < \epsilon$
\end{algorithmic}
\end{algorithm}
\vspace{-2.0mm}

Each iteration of Algorithm \ref{Algthm_AO} increases the objective function, which is bounded above for a given power constraint, until convergence.
The global optimality of the limit point cannot be guaranteed  due to non-convexity.
However, the stationarity (KKT optimality) of the solution can be argued based on the ideas in \cite{Razaviyayn2013}, 
avoided here due to space limitations.
\section{Simulation Results}
\label{Section_Simulations}
We consider i.i.d channels with entries drawn from $\mathcal{CN}(0,1)$, and  all results are averaged over $100$ channel realizations.
We compare the MMF rates for: 1) conventional beamforming (NoRS), 2) Single-Stream (SS) multigroup transmission described at the beginning of Section \ref{Section_RS}, and 3) the RS strategy.
The results for NoRS and SS are obtained using the SDR method in
\cite{Karipidis2008,Sidiropoulos2006}. We plot the SDR upper-bounds (no randomization), hence presenting optimistic performances for NoRS and SS.
On the other hand, the RS results and obtained by Algorithm \ref{Algthm_AO}, and represent the actual performances.
\begin{figure}
\vspace{-2.5mm}
\centering
\includegraphics[width = 0.4\textwidth]{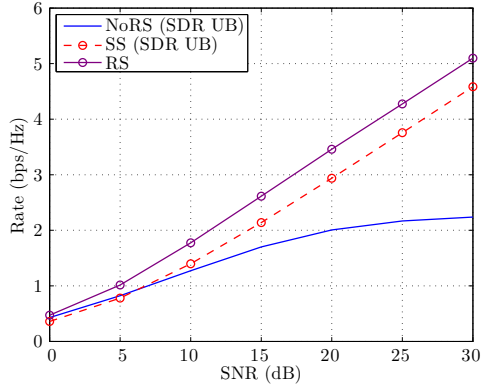}
\vspace{-3.0mm}
\caption{$N = 2$ antennas, $K = 4$ users, $M = 2$ equal groups.}
\label{Fig_1}
\vspace{-3.0mm}
\end{figure}
\begin{figure}
\vspace{-1.0mm}
\centering
\includegraphics[width = 0.4\textwidth]{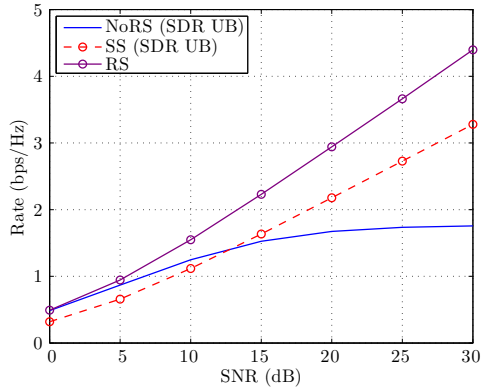}
\vspace{-3.0mm}
\caption{$N = 4$ antennas, $K = 9$ users, $M = 3$ equal groups.}
\label{Fig_2}
\vspace{-4.0mm}
\end{figure}

The MMF rates for a system with $N = 2$ transmit antennas and $M=2$
groups with $G=2$ users each are presented in Fig. \ref{Fig_1}.
As predicted from the DoF result in Proposition \ref{Proposition_DoF_NoRS}, NoRS exhibits a saturating performance.
Both SS and RS achieve non-saturating rates with DoFs of $1/2$ and an improved rate performance for RS which comes from  the designated beams.
The gains of RS over NoRS are very pronounced.
The results for a system with $N = 4$ transmit antennas and $M = 3$ groups with $G = 3$ users per group are shown in Fig. \ref{Fig_2}.
The benefits of using designated beams over SS transmission at low SNRs are clearer in this scenario, as the performance of the latter is constrained by the worst out of $9$ users.
While SS achieves a DoF of $1/3$, RS seems to surpass this DoF, which is evident from its slope at high SNR.
This suggests that the trivial achievable lower-bound in Proposition \ref{Proposition_DoF_RS} is in fact loose, and RS can achieve even higher MMF-DoF.
\section{Conclusion}
\label{Section_Conclusion}
In this paper, we proposed a RS multi-group multicast beamforming strategy.
We showed through DoF analysis that the proposed RS strategy outperforms the conventional beamforming strategy in overloaded scenarios, i.e. when the number of transmit antennas is insufficient to cope with inter-group interference.
An AO algorithm based on the WMMSE method was used to obtain the RS precoders.
The effectiveness of the proposed algorithm and the significant gains associated with the RS strategy were demonstrated through simulations.
Simulations also revealed that the trivial MMF-DoF lower-bound is in fact loose, which calls for a rigorous characterization of the optimum MMF-DoF achieved through RS.
\appendix[]
\begin{proof}[Proof of $d^{\ast} = 0$ for $N < N_{\min}$]
First, we write a precoding vector as $\mathbf{p}_{m} = \sqrt{q_{m}} \widehat{\mathbf{p}}_{m}$, where $q_{m}$ is the power and $\widehat{\mathbf{p}}_{m}$ is the unit-norm beamforming direction.
For a given precoding scheme characterized by one precoder for each power level, the $m$th power scales as $q_{m} = O(P^{a_{m}})$ with $a_{m} \leq 1$, further assumed to be non-negative as MMF necessitates non-vanishing powers allocated to all groups.
Let $\mathcal{I}_{m} \subset \mathcal{M}$ be the index set of groups that interfere with the $m$th group, depending on the precoder design.
From the DoF definitions in Section \ref{Subsection_DoF_definition}, it can be shown that
\begin{equation}
\label{Eq_DoF_UB}
d_{m} \leq \big( a_{m} - \max_{j \in \mathcal{I}_{m}} a_{j} \big)^{+}.
\end{equation}
The $(.)^{+}$ can be omitted as when it is active, the MMF-DoF is limited to zero which is also achieved when all $a_{m}$ are equal.
It is sufficient to show that the MMF-DoF is upper-bounded by $0$ for $N = N_{\min} - 1 = (M-1)G$, as decreasing the number of antennas does not increase the DoF.
For this case, $\mathbf{p}_{m}$ can be placed in the null space of at most $M-2$ groups, i.e. each beam interferes with at least one group.
It follows that $\bigcup_{m \in \mathcal{M}} \mathcal{I}_{m} = \mathcal{M}$.
We assume that each beam interferes with exactly one group, as the contrary does not increase the DoF.
It follows that at least two groups see non-zero interference.
Let $m_{1}$ be the index of the group receiving the dominant interference, i.e. $\max \{{a}_{m}\}_{m\in \mathcal{M}} \in \mathcal{I}_{m_{1}}$,
and $m_{2}$ be the index of the group receiving interference from $m_{1}$, i.e.  ${a}_{m_{1}} \in \mathcal{I}_{m_{2}}$.
It follows from \eqref{Eq_DoF_UB} that $d_{m_{1}} \leq a_{m_{1}} - a_{m_{2}}$ and $d_{m_{2}} \leq a_{m_{2}} - a_{m_{1}}$.
Since the symmetric-DoF is upper-bounded by the average of any number of group DoFs, we write
$d \leq \frac{d_{m_{1}} + d_{{m_{2}}}}{2} \leq 0 $, which holds for any possible precoder design.
\end{proof}

\ifCLASSOPTIONcaptionsoff
  \newpage
\fi

\bibliographystyle{IEEEtran}
\bibliography{IEEEabrv,References}


\end{document}